\documentclass[letterpaper, 10 pt, conference]{ieeeconf}  
\usepackage{amsmath,amsfonts}
\usepackage{algorithm2e}
\RestyleAlgo{ruled} 
\usepackage{array}
\usepackage[caption=true,font=footnotesize,labelfont=footnotesize,textfont=footnotesize]{subfig}
\usepackage{textcomp}
\usepackage{stfloats}
\usepackage{url}
\usepackage{verbatim}
\usepackage{graphicx}
\usepackage[dvipsnames]{xcolor}
\usepackage{cite} 
\usepackage{hyperref}
\usepackage{algpseudocode}
\usepackage{cases}
\usepackage{environ}

\newcommand{\proj}[2][1]{\text{proj}_{#1}(#2)}

\newtheorem{definition}{Definition}

\newtheorem{lemma}{Lemma}
\newtheorem{theorem}{Theorem}

\IEEEoverridecommandlockouts

\title{Threshold Strategy for Leaking Corner-Free Hamilton-Jacobi Reachability with Decomposed Computations}

\author{Chong He, Mugilan Mariappan, Keval Vora and Mo Chen
    \thanks{*This work was supported by the Canada CIFAR AI Chairs and NSERC Discovery Grants Programs.}
    \thanks{All authors are with the school of Computing Science, Simon Fraser University, Burnaby, BC V5A 1S6, Canada {\tt\small \{chong\_he, mmariapp, keval, mochen\}@sfu.ca}.}
}

\begin{document}

\maketitle

\begin{abstract}
Hamilton-Jacobi (HJ) Reachability is widely used to compute value functions for states satisfying specific control objectives.
However, it becomes intractable for high-dimensional problems due to the curse of dimensionality.
Dimensionality reduction approaches are essential for mitigating this challenge, whereas they could introduce the ``leaking corner issue", leading to inaccuracies in the results.
In this paper, we define the ``leaking corner issue" in terms of value functions, propose and prove a necessary condition for its occurrence. 
We then use these theoretical contributions to introduce a new local updating method that efficiently corrects inaccurate value functions while maintaining the computational efficiency of the dimensionality reduction approaches. 
We demonstrate the effectiveness of our method through numerical simulations.
Although we validate our method with the self-contained subsystem decomposition (SCSD), our approach is applicable to other dimensionality reduction techniques that introduce the ``leaking corners".
\end{abstract}

\section{Introduction}
Nonlinear systems control technologies have received growing interest in recent years and have many important applications~\cite{bansal2017hamilton}.
The applicable fields include legged robots ~\cite{wensing2023optimization}, safe collaboration between human and robotic systems~\cite{natarajan2023human}, and autonomous vehicles~\cite{hanover2024autonomous}, especially when including tasks ensuring safety by avoiding a failure set (safety task) and successfully reaching a goal set (liveness task). 
We generally call the failure sets and goal sets the target sets in our study.

Reachability analysis is generally used to ensure the safety and liveness of nonlinear systems.
Researchers have developed various methods to analyze reachability \cite{kong2015dreach, huang2019reachnn, mitchell2005time}, among which Hamilton-Jacobi (HJ) reachability excels at handling general nonlinear dynamics.
HJ reachability formulates an initial value function based on the signed distance function of a target (goal or obstacle) and updates it backward in time using optimal control. 
This value function provides insights into the robot’s capabilities for achieving tasks.
However, HJ reachability suffers from the curse of dimensionality. 
As the system’s dimensionality increases, computational complexity grows exponentially, making high-dimensional problems computationally infeasible.

Various approaches have been proposed to address the curse of dimensionality.
These include linearization of nonlinear dynamics \cite{spaceex2011frehse, althoff2010computing},  using hopf-lax formula \cite{darbon2016algorithms, lee2023efficient, sharpless2024conservative}, value function over- or under-approximation \cite{kurzhanski2000ellipsoidal,li2018approx,liu2024efficient}, reinforcement learning techniques \cite{fisac2019bridging, bansal2021deepreach, lin2024verification, ruthotto2020machine}, warm-starting with specific initializations \cite{herbert2019warm}, and paralleling the computation \cite{chow2019algorithm, bui2022optimizeddp}.
Additionally, methods for decoupling or reducing high-order dynamics have been explored \cite{mitchell2011scalable, holmes2020reachable}, including the method for composition into self-contained subsystems (SCSs) \cite{chen2017exact,chen2018decomposition}.

High-order dynamics decoupling methods are particularly advantageous due to their inherent parallelizability and immediate reduction of dimensionality.
Compared to other computational acceleration methods, they offer significant computational time reduction while preserving the key system properties.
It is achieved by computing the sub-value functions in low-dimensional subspaces and reconstructing the value function in full-dimensional space.
However, in certain problem formulations, the value function reconstructed in full-dimensional space will deviate from the value function directly computed from the full-dimensional system.
As a result, it can fail to guarantee liveness or safety for some states. This phenomenon is known as the ``leaking corner issue."~\cite{chen2018decomposition, lee2019removing}

A previous method for detecting the ``leaking corners" involves computing the admissible control set in low-dimensional subsystems and comparing them to identify affected states~\cite{he2023efficient}.
However, it is currently restricted to the self-contained subsystem decomposition (SCSD) method and only applies to cases with scalar control inputs.

In this paper, we provide:
\begin{itemize}
\item A formal definition of ``leaking corners" from the perspective of value functions.
\item A novel detection method capable of handling both scalar and vector control inputs without additional computation, which is also general to methods with decomposed computations.
\item A local updating procedure that ensures accurate results even without complete knowledge of all ``leaking corners," while maintaining computational efficiency.
\end{itemize}

\section{Background}

\subsection{System Dynamics}
Consider the following control-affine dynamical system:
\begin{equation}
\label{eq:dyn}
\frac{dz}{dt}=\dot{z}=f(z) + g(z)^\top u, \quad t\leq 0
\end{equation}

\noindent where $z\in\mathcal{Z}\subseteq\mathbb{R}^n$ denotes the system state within a state space, 
$u\in\mathcal{U}\subset\mathbb{R}^m$ 
denotes the control input.
We assume that the control space $\mathcal U$ is given by  the constraint 
\begin{equation}
\label{eq:control_constraint}
    c(u) = \| \alpha \odot u\|_\beta - \overline u \leq 0.
\end{equation}

\noindent where the weight
$\alpha\in\mathcal{A}\subseteq\mathbb{R}^m$, the norm $\beta\geq 1$, and the constraint value $\overline u \ge 0$.
The symbol ``$\odot$" represents the Hadamard product, which takes two matrices of the same dimensions as input and produces a matrix containing the multiplication of the corresponding elements in the inputs. 

We assume the control function $u(\cdot): [t,0] \mapsto \mathbb U$ is drawn from the set of measurable functions. 
We also assume that $f: \mathcal{Z} \mapsto \mathbb R^n$ and $g: \mathcal{Z} \mapsto \mathbb R^{n \times m}$ are such that the dynamics \eqref{eq:dyn} uniformly continuous, bounded, and Lipschitz continuous in $z$. 

\subsection{Value Functions in Hamilton-Jacobi Reachability}

Hamilton Jacobi (HJ) reachability analysis is an optimal control problem used to analyze the liveness and safety properties of dynamical systems. 
The cost function $\ell:\mathbb{R}^n\mapsto \mathbb{R}$ of the optimal control problem is designed such that its $0$ sub-level set is the goal (or failure) set, generally called the target set: $\mathcal{T}=\{z:\ell(z) \leq 0 \}$. 
A common choice of the cost function is the signed distance function to the set $\mathcal{T}$.

The value function is the solution of the terminal value HJB-PDE
\vspace{-2mm}
\begin{equation}
\label{eq:HJBPDE}
    D_t V(z,t)+ H(z,u,D_z V(z,t)) = 0
\vspace{-2mm}
\end{equation}

\noindent with Hamiltonian for liveness and safety cases
\begin{subequations} \label{eq:hamiltonian}
\begin{align}
    H_R(z,u,p) = \min_{u\in\mathcal{U}}\{p^\top f(z) + p^\top g(z)^\top u\} \label{eq:ham_reach}
    \\
    H_A(z,u,p) = \max_{u\in\mathcal{U}}\{p^\top f(z) + p^\top g(z)^\top u\} \label{eq:ham_avoid}
\end{align}
\end{subequations}

\noindent where $D_t$ and $D_z$ represent the derivative with respect to $t$ and $z$ respectively. 
The boundary condition is given by a final value function $ V(z,0)=\ell(z)$, and the value function $V(z,t)$ can be computed via applying dynamic programming backward in time until $t<0$.
Whenever necessary, we will add the subscript of $V$, and use $V_R(z,t)$ and $V_A(z,t)$ respectively to represent the value functions for liveness and safety problems.

The optimal control $u^*$ is obtained as follows:
\begin{subequations}\label{eq:optimal_control}
\begin{align}
u_R^* = \arg\min_{u\in\mathcal{U}} H_R(z,u,D_z V_R(z,t)), \\
u_A^* = \arg\max_{u\in\mathcal{U}} H_A(z,u,D_z V_A(z,t)).
\end{align}
\end{subequations}

We will omit the subscript of $u^*$ when convenient and use $u^*$ to denote the optimal control obtained from solving Eq.~\eqref{eq:optimal_control}.

\subsection{Computational Dimensionality Reduction} \label{subsec:dim_red}

Numerically, the HJB PDE~\eqref{eq:HJBPDE} is solved on a discrete grid; therefore, the computation scales exponentially with state dimension. 
This motivates the use of dimensionality reduction methods to reduce computational costs. 
Reducing computational dimensionality while maintaining exact global results is challenging.
The low-dimensional subsystems of \textit{coupled} nonlinear systems are defined below.

In this paper, for simplicity and clarity of presentation, we assume that there are two subsystems, but the results presented in the paper generalize to an arbitrary number of subsystems.

\begin{definition}
\label{def: Subsystem}
(Subsystem) Consider the special case in which the state $z$ can be expressed as $z=(z_1, z_2, z_c)$, with $z_1\in\mathbb{R}^{n_1}$, $z_2\in\mathbb{R}^{n_2}$, $z_c\in\mathbb{R}^{n_c}$, $n_1,n_2>0$, $n_c\ge 0$, and $n_1+n_2+n_c=n$. 
Following \cite{chen2018decomposition}, we call $z_1$, $z_2$, $z_c$  ``state partitions" of the system. 

Define $x_1 = (z_1, z_c)\in\mathcal{X}_1 \subseteq \mathbb{R}^{n_1+n_c}$ as the state of subsystem 1, and $x_2 = (z_2, z_c)\in\mathcal{X}_2 \subseteq \mathbb{R}^{n_2+n_c}$ as the state of subsystem 2.

We also express $u$ as $u=(u_1,u_2,u_c)$, with $u_1\in\mathbb{R}^{m_1}$, $u_2\in\mathbb{R}^{m_2}$, $u_c\in\mathbb{R}^{m_c}$, $m_1,m_2>0$, $m_c\geq 0$, and $m_1+m_2+m_c=m$.
We call $u_1$, $u_2$, and $u_c$ ``control partitions" of the system.

Let $w_1 = (u_1, u_c) \in \mathcal{W}_1 \subseteq \mathbb{R}^{m_1+m_c}$ be the control signal of subsystem 1, and $w_2 =(u_2, u_c) \in \mathcal{W}_2 \subseteq \mathbb{R}^{m_2+m_c}$ be the control signal of subsystem 2.

In the case of control-affine systems \eqref{eq:dyn}, we can write the dynamics of the two subsystems as 
\begin{subequations} \label{eq:subsys_dyn}
    \begin{align}
        \dot x_1 = f_1(x_1)+g_1(x_1)^\top w_1, \\
        \dot x_2 = f_2(x_2)+g_2(x_2)^\top w_2.
    \end{align}
\end{subequations}

The control constraints for each of the subsystems are as follows:
\vspace{-2mm}
\begin{subequations} \label{eq:subsystem_constraint}
    \begin{align}
        c_1(w_1) = \|a_1 \odot w_1\|_{\beta} - \overline u\leq 0,\label{eq:subsystem_constraint1} \\
        c_2(w_2) = \|a_2 \odot w_2\|_{\beta} - \overline u\leq 0. \label{eq:subsystem_constraint2}
        \vspace{-2mm}
    \end{align}
\end{subequations}
where $a_1=(\alpha_1,\alpha_c)$ and $a_2=(\alpha_2,\alpha_c)$.
\end{definition}

In terms of the low-dimensional controls, the constraint in Eq.~\eqref{eq:control_constraint} can be rewritten as 
\begin{equation} \label{eq:coupled_constraint}
     c_\text{joint}(w_1, w_2) \le 0,
\end{equation}

\noindent where $c_\text{joint}(w_1, w_2) = c(u)$ can be expanded as follows:
\begin{align} 
    &c_\text{joint}(w_1, w_2) = \|[\alpha_1, \alpha_2, \alpha_c] \odot [u_1, u_2, u_c]\|_\beta - \overline u 
    \\ 
    &=\sqrt[\uproot{5}\beta]{\sum_{j=1}^{m_1}|\alpha_{1,j} u_{1,j}|^\beta+\sum_{j=1}^{m_2}|\alpha_{2,j} u_{2,j}|^\beta + \sum_{j=1}^{m_c}|\alpha_{c,j} u_{c,j}|^\beta} - \overline u \nonumber
\vspace{-2mm}
\end{align}

The projection operator shows how the full-dimensional state $z$ is related to the low-dimensional state $x_i$.
\begin{definition}
\label{def:proj}
(Projection operator)
    A projection operator $\text{proj}: \mathbb R^n \mapsto \mathbb R^{n_i+n_c}$ maps a state $z$ in full-dimensional space $\mathbb R^n$ to a state $x_i$ in low-dimensional subspace $\mathbb R^{n_i+n_c}$:
    \vspace{-2mm}
    \begin{equation}
        \proj[i]{z} := (z_i,z_c) = x_i.
    \end{equation}
\end{definition}

The ``leaking corner issue" also occurs without the dimensional reduction approaches.
In thoses cases, both equations in Eq.~\eqref{eq:subsys_dyn} become the full-dimensional system in Eq.~\eqref{eq:dyn}, and our method will still work.

\textit{Running example:} 
Consider decomposing the 2D Single Integrator Model $z = (p_x, p_y)$ using the method in~\cite{chen2018decomposition}:
\vspace{-2mm}
\begin{equation}
\label{eq:2D_SI_full}
    \dot p_x = u_x \quad
    \dot p_y = u_y ,
\vspace{-2mm}
\end{equation} 
The state partitions of the system are $p_x$ and $p_y$. 
The control input is $u = (u_x, u_y)$  and constrained by $\|u\|_2 \leq \overline u$. 

There are 2 low-dimensional subsystems, and the two subsystems are
\vspace{-2mm}
\begin{equation}
\label{eq:2D_SI_low}
\begin{aligned}
    \text{Subsystem 1:  } x_1 = p_x, \quad
    \text{Subsystem 2:  } x_2 = p_y. 
\end{aligned}
\vspace{-2mm}
\end{equation}

Subsystem $1$ has a control input given by $w_1 = u_x$, while subsystem $2$ has $w_2=u_y$.
The control constraints in low-dimensional systems are as:
\vspace{-1mm}
\begin{equation}
\begin{aligned}
    &\text{Subsystem 1:  } c_1(w_1)=\|u_x\|_2 \leq \overline u, \\
    &\text{Subsystem 2:  } c_2(w_2)=\|u_y\|_2 \leq \overline u 
\end{aligned}
\vspace{-1mm}
\end{equation}
to best preserve the information.           

\subsection{Value Function Reconstruction} \label{subsec:comb_vf}

When applying the computational acceleration method, the low-dimensional value function is computed through the subsystem dynamics as in Definition~\ref{def: Subsystem}.
Consequently, the initial value function representing the target set also needs to be modified for low-dimensional computation.
The cost function associated with the low-dimensional computation is given by $\ell_i:\mathbb{R}^{n_i+n_c} \mapsto \mathbb{R}$.

The low-dimensional value function is the solution of the terminal value HJB-PDE
\vspace{-1mm}
\begin{equation} \label{eq:low_HJB}
    D_t \phi_i(x_i,t)+H_{\phi_i}(x_i,w_i,D_{x_i} \phi_i(x_i,t))=0
    \vspace{-1mm}
\end{equation}
with the boundary condition being given by $\phi_i(x_i,0)=\ell_i(x_i)$.
The low-dimensional optimal control $w_i^*$ is obtained as follows:
\vspace{-2mm}
\begin{subequations}\label{eq:low_optimal_control}
\begin{align}
w_{R,i}^*(t) = \arg\min_{w_i\in\mathcal{W}_i}  H_{\phi_i}(x_i,w_i,D_{x_i}\phi_i(x_i,t)), \\
w_{A,i}^*(t) = \arg\max_{w_i\in\mathcal{W}_i}  H_{\phi_i}(x_i,w_i,D_{x_i}\phi_i(x_i,t)).
\end{align}
\vspace{-3mm}
\end{subequations}

\begin{definition} 
\label{def:subvalue_func}
(Full-dimensional sub-value function)
The full-dimensional sub-value function $V_i(z,t)$ is defined as the value function of the subsystem $i$ with respect to the full-dimensional space $\mathbb{R}^n$.
\vspace{-1mm}
\begin{equation}\label{eq:full_subvalue_func}
    V_i(z,t) = \phi_i(x_i,t), \text{ where } x_i=\proj[i]{z}
\end{equation}
\end{definition}
$V_i(z,t)$ could be different depending on the method.\footnote[1]
{For Mixed Implicit Explicit (MIE) formulation~\cite{mitchell2011scalable}, the full-dimensional sub-value functions are
$V_i(z,t) = \gamma_i(\phi_i(x_1,t), x_2)$, where $x_1=\proj[1]{z}$ and $x_2=\proj[2]{z}$.
$\gamma_i: \mathbb{R}\times \mathbb{R}^{n_2+n_c}\mapsto \mathbb{R}$ is continuous.
}

\subsubsection{}
The compuation is termed an \textbf{intersection} case when the full-dimensional initial value function relates to the full-dimensional initial sub-value functions as
\vspace{-1mm}
\begin{equation}
\label{eq:intersection}
V(z,0)=\max_i (V_i(z,0)).
\vspace{-2mm}
\end{equation}

\noindent The point-wise maximum operation corresponds to the intersection of sets, following the convention of representing sets using sublevel sets of functions \cite{mitchell2007toolbox}.

\subsubsection{}
The compuation is termed \textbf{union} case when the full-dimensional initial value function is related to the full-dimensional initial sub-value functions as
\vspace{-2mm}
\begin{equation}
\label{eq:union}
V(z,0)=\min_i (V_i(z,0)).
\vspace{-2mm}
\end{equation}

\begin{definition}\label{def: aprox_valuef}
    (Approximated value function)
    Given the full-dimensional sub-value functions $V_i(z,t)$, the approximated (full-dimensional) value function $\hat{V}(z,t)$ are given, based on the \textbf{intersection} or \textbf{union} cases, as follows:
    \vspace{-2mm}
    \begin{subequations}\label{eq:approx_v} 
    \begin{align}
    \textbf{Intersection}: \hat{V}(z,t)=\max_i \{ V_i(z,t)\};  \label{eq:approx_intersect}
    \\
    \textbf{Union}:\quad \hat{V}(z,t)=\min_i \{V_i(z,t)\}.  \label{eq:approx_union}
    \end{align}
    \end{subequations}
\end{definition}

The approximated value functions sometimes provide safety and liveness guarantees with better computational efficiency.
However, they also sometimes vary from the directly computed true value function.
We refer to the latter phenomenon as the ``leaking corner issue"~\cite{chen2017exact,chen2018decomposition,lee2019removing}.

\section{leaking corners}

\begin{definition}
\label{def: leakingcorner}
    (Leaking Corners)
    Suppose we obtain $V(z,t)$ by solving the HJ PDE \eqref{eq:HJBPDE}, $\hat{V}(z,t)$ via  Eq.~\eqref{eq:approx_v}.  
    The set of ``leaking corners" $\mathcal{L}(t)$ is defined as
    \vspace{-2mm}
    \begin{equation}
        \mathcal{L}(t) = \{z: V(z,t)\neq \hat{V}(z,t)\}.
    \end{equation}
\end{definition}

Based on Theorem 1 and 2 in~\cite{chen2018decomposition}, there are $2$ cases that will have the ``leaking corners" :

\noindent(1) The intersection case for the liveness problem, where we obtain $\hat V_R$ as follows: 
\vspace{-2mm}
\begin{equation}
\label{eq:LC_reach_intersect}
    \hat V_R(z,t) = \max\{V_{R,i}(z, t)\};
    \vspace{-2mm}
\end{equation}
(2) The union case for the safety problem:
\vspace{-2mm}
\begin{equation}
\label{eq:LC_avoid_union}
    \hat V_A(z,t)=\min\{V_{A,i}(z,t)\}.
\vspace{-2mm}
\end{equation}

Intuitively, the ``leaking corner issue" arises due to the mismatch between the control inputs in different low-dimensional subsystems.
The following subsections formalize this intuition.

\subsection{Allowable Control and the ``Leaking Corner"}

The low-dimensional value functions could evolve independently.
However, due to the coupled constraints on the low-dimensional controls, the optimal controls~\eqref{eq:low_optimal_control} may not always satisfy the original constraints \eqref{eq:coupled_constraint}.
To address this, we introduce the concept of allowable control.

\begin{definition}\label{def:allow_control}
    (Allowable Control)  
    We define a pair of low-dimensional control signals $(\tilde w_1, \tilde w_2)$ that satisfies the coupled constraint in Eq.~\eqref{eq:coupled_constraint} as \textit{allowable controls}.
    We also define \textit{allowable control functions} $(\tilde w_1(\cdot), \tilde w_2(\cdot))$ as control functions that satisfy Eq.~\eqref{eq:coupled_constraint} for all time.
    
    The corresponding full-dimensional sub-value function corresponding to allowable control functions $(\tilde w_1(\cdot), \tilde w_2(\cdot))$ is denoted $\tilde V_i(z,t)$.

    For convenience, assume $\tilde w_1$  is the optimal control $w^*_1$ given in Eq.\eqref{eq:low_optimal_control}, we will denote the other component of the pair of \textit{allowable controls} as $\tilde w^*_2$ with $c_\text{joint}(w^*_1, \tilde w^*_2)\le 0$. The corresponding value functions are denoted as $V_1(z,t)$ and $\tilde V^*_2(z,t)$.
\end{definition}

\begin{lemma} \label{lem:coupled_control_0}
    Suppose $\tilde w_1(t) = w_1^*(t):=(u_1(t), u_c(t))$ for all $t$ and $(\tilde w_1(\cdot), \tilde w_2(\cdot))$ is a pair of allowable control functions. 
    Then, $\tilde w_2(t)=\tilde w_2^*(t)=(u_2(t), u_c(t))$, where $ u_2(t)=\mathbf{0}$ for all $t$.
\end{lemma}

\begin{proof}
    First, note that $\|\cdot\|_\beta$ is convex and the objectives in Eq.~\eqref{eq:low_optimal_control} are linear in $w_i$.
    This means constraint in Eq.~\eqref{eq:subsystem_constraint1} at the optimal control $w_1^*$ of subsystem 1 must be tight:

    \begin{equation}
        c_1(w_1^*) = \|a_1 \odot w_1^* \|_\beta - \overline u =0
    \end{equation}

    Next, since the joint control constraints in Eq.~\eqref{eq:coupled_constraint} are given by $0 \ge c_\text{joint}(w_1^*,\tilde w^*_2)$, we have the following:
    
    {\small
    \begin{subequations} \label{eq:lem_1_proof_3}
        \begin{align}
            0 &\ge c_\text{joint}(w_1^*,\tilde w^*_2) \\
            & =\sqrt[\uproot{5}\beta]{\sum_{j=1}^{m_1}|\alpha_{1,j} u_{1,j}|^\beta+\sum_{j=1}^{m_2}|\alpha_{2,j} u_{2,j}|^\beta + \sum_{j=1}^{m_c}|\alpha_{c,j} u_{c,j}|^\beta} - \overline u \label{eq:bigger_root} \\
            &\geq \sqrt[\uproot{5}\beta]{\sum_{j=1}^{m_1}|\alpha_{1,j} u_{1,j}|^\beta+\sum_{j=1}^{m_c}|\alpha_{c,j} u_{c,j}|^\beta} - \overline u \label{eq:smaller_root} \\
            &= \|a_1 \odot w_1^*\|_\beta  - \overline u = 0
        \end{align}
    \end{subequations}
    }

    Thus, $c_\text{joint}(w_1^*,\tilde w^*_2)=0$;
    in particular, this means that the expressions in \eqref{eq:bigger_root} and \eqref{eq:smaller_root} are equal, which implies $\sum_{j=1}^{m_2}|\alpha_{2,j} u_{2,j}|^\beta=0$.
    This is equiavlent to $u_2=\mathbf{0}$.
    
    In addition, this must be true at every time $t$. 
    Therefore, we have that $\tilde w^*_2(t)=(u_2(t), u_c(t))$, where $ u_2(t)=\mathbf{0}$ for all $t$. 
\end{proof}

\begin{lemma}\label{lem:allow_control_no_LC}
    A state $z$ is not in the leaking corner, $z \notin \mathcal L(t)$, if and only if there exists a pair of allowable control functions $\tilde w_1 (\cdot)$ and $\tilde w_2(\cdot)$ such that they satisfy the following: 
    \vspace{-2mm}
    \begin{subequations}
    \label{eq:lem_2_condition}
    \begin{align}
    &\text{Case in Eq.~\eqref{eq:LC_reach_intersect}: }
    \max \{\tilde V_{R,1}, \tilde V_{R,2}\}=\hat{V}_R, \label{eq:allow_no_LC_reach}
    \\
    &\text{Case in Eq.~\eqref{eq:LC_avoid_union}: }
    \min \{\tilde V_{A,1}, \tilde V_{A,2}\}=\hat{V}_A. \label{eq:allow_no_LC_avoid}
    \end{align}
    \end{subequations}
\end{lemma}

\begin{proof}
    We will begin with proving the intersection case for liveness problem (Eq.~\eqref{eq:LC_reach_intersect}):
    
    Theorem 2 in~\cite{he2023efficient} implies, in terms of value functions, that
    \vspace{-2mm}
    \begin{equation}
    \label{eq:lem_2_proof_1}
        \max \{\tilde V_{R,1}(z,t), \tilde V_{R,2}(z,t)\} \geq V_R(z,t).
        \vspace{-2mm}
    \end{equation}
    Additionally, from Lemma 2 in~\cite{he2023efficient}, we know that 
    \vspace{-2mm}
    \begin{equation}
    \label{eq:lem_2_proof_2}
        V_R(z,t) \geq \hat{V}_R(z,t).
        \vspace{-2mm}
    \end{equation}
    Combining Eqs.~\eqref{eq:lem_2_proof_1} and \eqref{eq:lem_2_proof_2}, we obtain
    \vspace{-2mm}
    \begin{equation}
        \max \{\tilde V_{R,1}, \tilde V_{R,2}\} \geq V_R \geq \hat{V}_R.
    \vspace{-2mm}
    \end{equation}
    
    Thus, if $\max \{\tilde V_{R,1}(z,t), \tilde V_{R,2}(z,t)\}=\hat{V}_R(z,t)$, we have $V_R(z,t) = \hat V_R(z,t)$, which is equivalent to $z\notin \mathcal L(t)$.   

    Conversely, if $z\notin \mathcal L$(t), then $V_R(z,t) = \hat V_R(z,t)$. 
    By Theorem 1 in~\cite{he2023efficient}, there exists a pair of allowable control functions ensuring $\max \{\tilde V_{R,1}(z,t), \tilde V_{R,2}(z,t)\} = V_R(z,t)$.

    The proof for the case described in Eq.~\eqref{eq:LC_avoid_union} follows the same reasoning and is omitted for brevity.  
\end{proof}

\section{Correcting of Leaking Corners}
In this section, we present a method for detecting and correcting the ``leaking corners" using full-dimensional sub-value functions. 

\subsection{Detecting the Leaking Corners}

Based on the allowable controls and their corresponding value functions in Definition~\ref{def:allow_control}, we can detect the leaking corners $\mathcal{L}(t)$ by value comparison.

\begin{theorem}\label{thrm:leaking_corner_delta} 

    We can find the set of leaking corners $\mathcal{L}(t)$ by comparing the (full-dimensional) sub-value functions.
    \begin{equation}
        \mathcal{L}(t) = \{z: |V_1(z,t) - V_2(z,t)| < \Delta\}.
    \end{equation}

    The value of $\Delta$ is
    \begin{subnumcases}{\Delta =}
    | \tilde V_{1}^* - V_{1} |, & \text{if $V_{R,2} \ge V_{R,1}$ or $V_{A,1}\ge V_{A,2}$}.\\
    | \tilde V_{2}^* - V_{2} |, & \text{if $V_{R,1} \ge V_{R,2}$ or $V_{A,2}\ge V_{A,1}$}.
    \end{subnumcases}

\end{theorem}

\begin{proof}
Without loss of generality, we will consider the liveness problem, in the case where $V_{R,1}(z,t)\ge V_{R,2}(z,t)$.
The case in which $V_{R,2}(z,t)\ge V_{R,1}(z,t)$ can be proven by repeating the exact same argument.
The cases involvign the safety problem can be proven by starting with $\min\{V_{A,1}(z,t), V_{A,2}(z,t)\}=\hat V_A(z,t)$ and $\tilde V_{A,1}(z,t) \le V_{A,1}(z,t)$, and then following the exact same arguments.

In the liveness problem, as stated in \eqref{eq:LC_reach_intersect}, we have $\max\{V_{R,1}(z,t), V_{R,2}(z,t)\}=\hat{V}_R(z,t)$. 
In the case that $V_{R,1}(z,t)\ge V_{R,2}(z,t)$, we have $V_{R,1}(z,t)=\hat V_R(z,t)$.

Note that $V_{R,1}(z,t) \le\tilde V_{R,1}(z,t)$ as it is solved by minimizing the Hamiltonian ~\eqref{eq:ham_reach}.
From Eq. \eqref{eq:allow_no_LC_reach} in Lemma~\ref{lem:allow_control_no_LC}, there exists a pair of allowable controls to let $\max\{\tilde V_{R,1}(z,t), \tilde V_{R,2}(z,t)\}=\hat{V}_R(z,t)$ if $z\notin\mathcal{L}(t)$.
Thus, $\tilde V_{R,1}(z,t)=V_{R,1}(z,t)=\hat{V}_R(z,t)$.

When $V_{R,1}\ge V_{R,2}$,
\begin{align}
    z\notin \mathcal{L}(t)& \Leftrightarrow  \tilde V_{R,1}=V_{R,1}=\hat{V}_R \wedge \max\{\tilde V_{R,1}, \tilde V_{R,2}\} = \hat{V}_R 
    \\
    &(\tilde w_1 = w_1^* \text{, and } \tilde w^*_2 \text{ can be found with Lemma~\ref{lem:coupled_control_0}} )\nonumber
    \\
    &\Leftrightarrow \tilde V^*_{R,2}=\tilde V_{R,2}\le \tilde V_{R,1} = V_{R,1}=\hat{V}_R \label{eq:proof_contradict}
\end{align}
Note that $V_{R,2}\le \tilde V_{R,2}$ as it is solved by minimizing the Hamiltonian ~\eqref{eq:ham_reach}, we define
\begin{equation}
    \Delta=|\tilde V^*_{R,2}-V_{R,2}| = \tilde V^*_{R,2}-V_{R,2},
\end{equation}
and let
\begin{align}
    |V_{R,1}-V_{R,2}| = V_{R,1}-V_{R,2}  &< \Delta
    \\
    V_{R,1}-V_{R,2} &< \tilde V^*_{R,2} - V_{R,2}
    \\
    V_{R,1} &< \tilde V^*_{R,2}
\end{align}

\noindent which contradicts Eq.~\eqref{eq:proof_contradict}, thus we prove that when $V_{R,1}(z,t)-V_{R,2}(z,t)<\Delta$, state $z\in\mathcal{L}(t)$.

\end{proof}

\subsection{Local Updating Procedure}

\begin{definition}
    (Island)
    The ``leaking corner" set $\mathcal{L}(t)$ consists of a finite union of $k$ connected sets, referred to as islands, denoted by $\mathcal{I}_\kappa(t)$ for $\kappa\in\{1,2,3, ..., k\}$. Each $\mathcal{I}_\kappa(t)$ is a connected component of $\mathcal{L}(t)$. 
\end{definition}

Assume that for all $\kappa\in\{1,2,3, ..., k\}$, there exists at least one state $z^*\in\mathcal{I}_\kappa(t)$ such that 
\vspace{-1mm}
\begin{equation}
\label{eq:IV_B_1}
    V_1(z^*,t) = V_2(z^*,t)
\vspace{-1mm}
\end{equation}
for both cases suffering from the ``leaking corner issue," as described in Eq.~\eqref{eq:LC_reach_intersect} and Eq.~\eqref{eq:LC_avoid_union}.

Since $\hat V(z^*,t)$ is obtained through Eq.~\eqref{eq:approx_v}.
If Eq.~\eqref{eq:lem_2_condition} and Eq.~\eqref{eq:IV_B_1} hold, the following condition must hold:
\vspace{-1mm}
\begin{equation}
\label{eq:IV_B_2}
    \tilde V_{1}(z^*,t) = V_{1}(z^*,t)\text{, and }\tilde V_{2}(z^*,t) = V_{2}(z^*,t).
\vspace{-1mm}
\end{equation}
However, to satisfy the above equation, optimal control in both subsystems must be applied, which violates the control constraint given by Eq.~\eqref{eq:coupled_constraint}.
The coupled control constraint attains its worst-case violation:
\vspace{-1mm}
\begin{equation}
    c(w_1^*, w_2^*)= \max_{w_1\in\mathcal{W}_1, w_2\in\mathcal{W}_2} c(w_1,w_2)>0.
\vspace{-1mm}
\end{equation}
Such a state $z^*$ is a  ``leaking corner" as indicated in Lemma~\ref{lem:allow_control_no_LC}.

Let $V_d(z,t)$ represent the absolute difference between the low-dimensional value functions:
\vspace{-1mm}
\begin{equation}
V_d(z,t) = | V_1(z,t) - V_2(z,t) |\geq 0.
\vspace{-1mm}
\end{equation}
$V_i(z,t)$ as Eq.~\eqref{eq:full_subvalue_func} is from the viscosity solution of a continuous initial value function \cite{evans2022partial}, then $V_i(z,t)$ and $V_d(z,t)$ are also continuous.

\begin{algorithm}[h]
\caption{Local updating procedure}\label{alg: local_update}
\KwData{$\hat{V}(\cdot, \cdot), \hat{\mathcal{L}}(\cdot), Z, t_\text{list}=[t, t+\delta ..., 0]$}
\KwResult{$\check{V}(\cdot, \cdot)$}
\SetKwProg{Def}{def}{:}{}

$s \gets 0$; \Comment{Backward Computation} 

$\check{V}(\cdot,0) \gets \hat{V}(\cdot, 0)$; \\
$\text{Frontier} \gets \text{nextFrontier} \gets \text{visited} \gets \{ \};$ 

\While{$s > t$}{
    \For{$z \in Z$}{
      \eIf{$z \in \hat{\mathcal{L}}(s)$}{
        $\text{updateValue}(z, s, \delta, \text{Frontier})$
      }{
          $\check{V}(z, s-\delta) \gets \hat{V}(z,s-\delta)$\;
      }
    }
    $\text{visited} \gets \hat{\mathcal{L}}(s)$ \\
    $\text{Frontier} \gets \text{Frontier} \setminus \text{visited}$ \\
    \While{$\text{Frontier} \neq $ \O}{         
        \For{$z$ in $\text{Frontier}$}{
            $\text{updateValue}(z, s, \delta, \text{nextFrontier})$        
        }
        $\text{visited} \gets \text{visited}\ \cup \ \text{Frontier} $\\
        $\text{Frontier} \gets \text{nextFrontier} \setminus \text{visited} $\\
        $\text{nextFrontier} \gets \{ \}$ 
    }
    $s \gets s - \delta$\;
} 

\Def{updateValue($z, s, \delta, \text{Frontier}$)}
{
    $\check{V}(z, s-\delta) \gets \text{HJ Update}(\check{V}(z, s) )$ \Comment{Equation ~\ref{eq:HJBPDE}} \\
    \If{$\check{V}(z, s-\delta) \neq \hat{V}(z, s-\delta)$}
    {
        $\text{Frontier} \gets \text{Frontier} \cup  \text{neighbor}(z)$ 
    }
}
\end{algorithm}

For neighboring states where $V_d(z,t)\geq 0$, there exist control options $(w_1,w_2)$ that are different from the optimal control pair $(w_1^*,w_2^*)$ if try to satisfy the condition given by Eq.~\eqref{eq:lem_2_condition}. 
The constraint violation is reduced, though it may not be eliminated. Mathematically, this is expressed as:
\vspace{-1mm}
\begin{equation}
    0\leq c(w_1,w_2)<c(w_1^*,w_2^*).
\vspace{-1mm}
\end{equation}
As we move slightly away from the worst-case mismatch points, we eventually reach states where allowable control pairs $(\tilde w_1, \tilde w_2)$ exist such that the constraint is satisfied:
\vspace{-1mm}
\begin{equation}
    c(\tilde w_1,\tilde w_2)\leq 0.
\vspace{-1mm}
\end{equation}
These states are not ``leaking corners" according to Lemma~\ref{lem:allow_control_no_LC}.

Since each connected island $\mathcal I_\kappa$ contains at least one worst-case scenario state $z_\kappa$, for each $\mathcal I_\kappa(t)$, there exists a state $z\in\mathcal{L}(t)$ from Theorem~\ref{thrm:leaking_corner_delta}.
The transition from extreme constraint violation to a region free from the ``leaking corner issue" occurs continuously within each island.

\begin{figure}[h]
{\includegraphics[width=\linewidth]{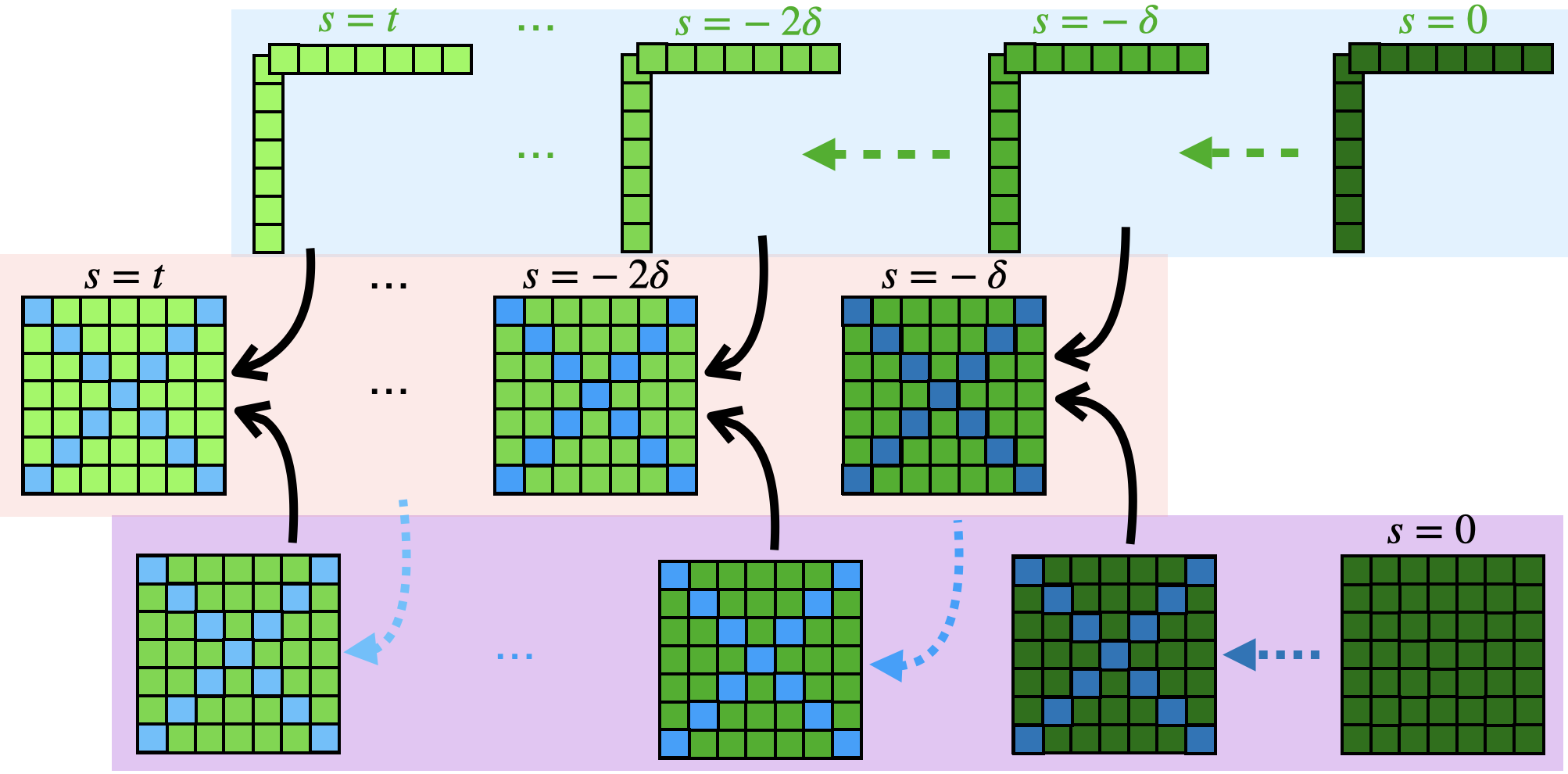}}
\vspace{-5mm}
\caption{The results from the low-dimensional computation are in the first row.
The third row shows the local updated results in the full-dimensional space.
The combined results, which equal the true results, are in the second row.
}
\label{fig: localprocedure}
\vspace{-3mm}
\end{figure}

To correct these ``leaking corners", we apply Algorithm~\ref{alg: local_update}, which locally updates the affected regions.
We first detect an approximated ``leaking corners" $\hat{\mathcal{L}}(\cdot)\subseteq\mathcal{L}(\cdot)$ through Theorem \ref{thrm:leaking_corner_delta}.
Using the approximated value function $\hat{V}(\cdot, \cdot)$ from Definition~\ref{def: aprox_valuef}, we locally refine the value of the ``leaking corners" through Algorithm~\ref{alg: local_update} and denote the updated result as $\check{V}(\cdot,\cdot)$.
Fig.~\ref{fig: localprocedure} provides a visual representation of this update procedure, illustrating how the detected leaking corners are iteratively corrected.

In Algorithm~\ref{alg: local_update}, $\text{neighbor}(z)$ returns a set of states that are adjacent to the state $z$.
Every island of the ``leaking corners" will be included, and every ``leaking corner" is covered using our formula.

\section{Numerical Examples}

In this section, the 2D and 6D examples demonstrate that:

\noindent(1) Theorem~\ref{thrm:leaking_corner_delta} accurately locates the ``leaking corners."

\noindent(2) The local updating process in Algorithm~\ref{alg: local_update} produces results equivalent to the ground truth while maintaining computational efficiency.

The experiment was conducted on a system with 96.0 GB of memory, an AMD Ryzen 9 5950X 16-core processor (32 threads), and Ubuntu 22.04.3 LTS as the operating system.
We use the highly parallelized Optimized\_dp~\cite{bui2022optimizeddp} for computation, while our correction step, implemented in Python, is not parallelized.
We choose the self-contained subsystem decomposition (SCSD) as the dimensionality reduction method for our experiments.

We consider the two values the same if their difference is within a threshold of $1 \times 10^{-3}$.

\subsection{Running Example: 2D Single Integrator}

Consider the 2D single integrator example, which is used as a demonstration throughout the paper.
Let the control constraint value be $\overline u = 1 \text{m/s}$.

The initial value function for subsystem 1 is $\phi_1(x_1,0)=\ell_1(x_1)=|x|-1$, and for subsystem 2, it is $\phi_2(x_2,0)=\ell_2(x_2)=|y|-1$.
The initial value function for the full-dimensional system is given by $V(z,0)=\max \{\ell_1(x_1), \ell_2(x_2) \}$.

The grid consists of $101\times 101$ points with each dimension ranging from $-4$ to $4$.

\subsubsection{Single time step case}

In this case, we set $t = -0.02 \text{s}$ and $\delta = 0.02 \text{s}$.
The value function updated with our method is denoted as $\check{V}(z,t)$.
For comparison, the ground truth value function $V(z,t)$ is obtained through direct computation from the full-dimensional system.

First, we identify the ``leaking corners" using Theorem~\ref{thrm:leaking_corner_delta}. 
The ground truth number of ``leaking corners" is $200$, and the detection using Theorem~\ref{thrm:leaking_corner_delta} with 
$\Delta=0.02$ identified $201$ ``leaking corners".
After locating the ``leaking corners", we perform the local updating process as outlined in Algorithm~\ref{alg: local_update}.

\begin{table}[h]
\vspace{-2mm}
    \centering
    \caption{2D Accuracy Comparison for One Step}
    \label{table: 2D_OneStep_Accuracy}
    \begin{tabular}{|p{3cm}|>{\centering\arraybackslash}p{2cm}|>{\centering\arraybackslash}p{2cm}|}
        \hline
        \textbf{Metric} & \textbf{Before} & \textbf{After} \\
        \hline
        Number of grid points with different values from the ground truth & 200 & 0 \\
        \hline
        Average absolute difference from ground truth & $1.2 \times 10^{-4}$ & $9.51 \times 10^{-18}$ \\
        \hline
        Maximum absolute difference from ground truth & $2 \times 10^{-2}$ & $2.22 \times 10^{-16}$ \\
        \hline
    \end{tabular}
\vspace{-2mm}
\end{table}

\begin{table}[h]
    \centering
    \caption{2D Time Comparison for One Step}
    \label{table: 2D_OneStep_Time}
    \begin{tabular}{|p{3cm}|>{\centering\arraybackslash}p{4cm}|}
        \hline
        \textbf{Process} & \textbf{Time (seconds)} \\
        \hline
        Direct computation & $3.3 \times 10^{-2}$ \\
        \hline
        SCSD computation + HJ local update computation & $7 \times 10^{-4} + 1.3 \times 10^{-3} = 2.0 \times 10^{-3}$ \\
        \hline
    \end{tabular}
    \vspace{-3mm}
\end{table}

Table~\ref{table: 2D_OneStep_Accuracy} presents the accuracy comparison, while Table~\ref{table: 2D_OneStep_Time} shows the computational time comparison. 
The results demonstrate that our method accurately locates and corrects the ``leaking corners." 
Additionally, compared to direct computation, our method offers better time efficiency.

\subsubsection{10 time steps case}

In this case, $t=-0.2\text{s}$ and $\delta=0.02\text{s}$. 
The computation needs to run backward for 10 time steps to complete.

For the final time step of computation, the ground truth number of the ``leaking corners" is $1344$. 
Using Theorem~\ref{thrm:leaking_corner_delta} with $\Delta =0.2$, we identified $1001$ ``leaking corners".
After detecting these ``leaking corners", we apply the local updating process described in Algorithm~\ref{alg: local_update}.

\begin{table}[h]
\vspace{-2mm}
    \centering
    \caption{2D Accuracy Comparison for 10 Steps}
    \label{table: 2D_10Step_Accuracy}
    \begin{tabular}{|p{3cm}|>{\centering\arraybackslash}p{2cm}|>{\centering\arraybackslash}p{2cm}|}
        \hline
        \textbf{Metric} & \textbf{Before} & \textbf{After} \\
        \hline
        Number of states with different values from the ground truth & 1344 & 0 \\
        \hline
        Average absolute difference from ground truth & $2.5 \times 10^{-3}$ & $1 \times 10^{-9}$ \\
        \hline
        Maximum absolute difference from ground truth & $7.39 \times 10^{-2}$ & $2.44 \times 10^{-8}$ \\
        \hline
    \end{tabular}
    \vspace{-5mm}
\end{table}

\begin{table}[h]
    \centering
    \caption{2D Time Comparison for 10 Steps}
    \label{table: 2D_10Step_Time}
    \begin{tabular}{|p{3cm}|>{\centering\arraybackslash}p{4cm}|}
        \hline
        \textbf{Process} & \textbf{Time (seconds)} \\
        \hline
        Direct computation & $3.36 \times 10^{-1}$ \\
        \hline
        SCSD computation + HJ local update computation & $4.72 \times 10^{-3} + 1.61 \times 10^{-1} = 1.65 \times 10^{-1}$ \\
        \hline
    \end{tabular}
    \vspace{-2mm}
\end{table}

Table~\ref{table: 2D_10Step_Accuracy} presents the accuracy comparison, while Table~\ref{table: 2D_10Step_Time} provides the computational time comparison.
The results show that our method successfully identifies every island of ``leaking corners."
Moreover, our approach improves time efficiency compared to direct computation, even without implementing parallel computing for making the corrections.

\begin{figure}[t]
{\includegraphics[width=\linewidth]{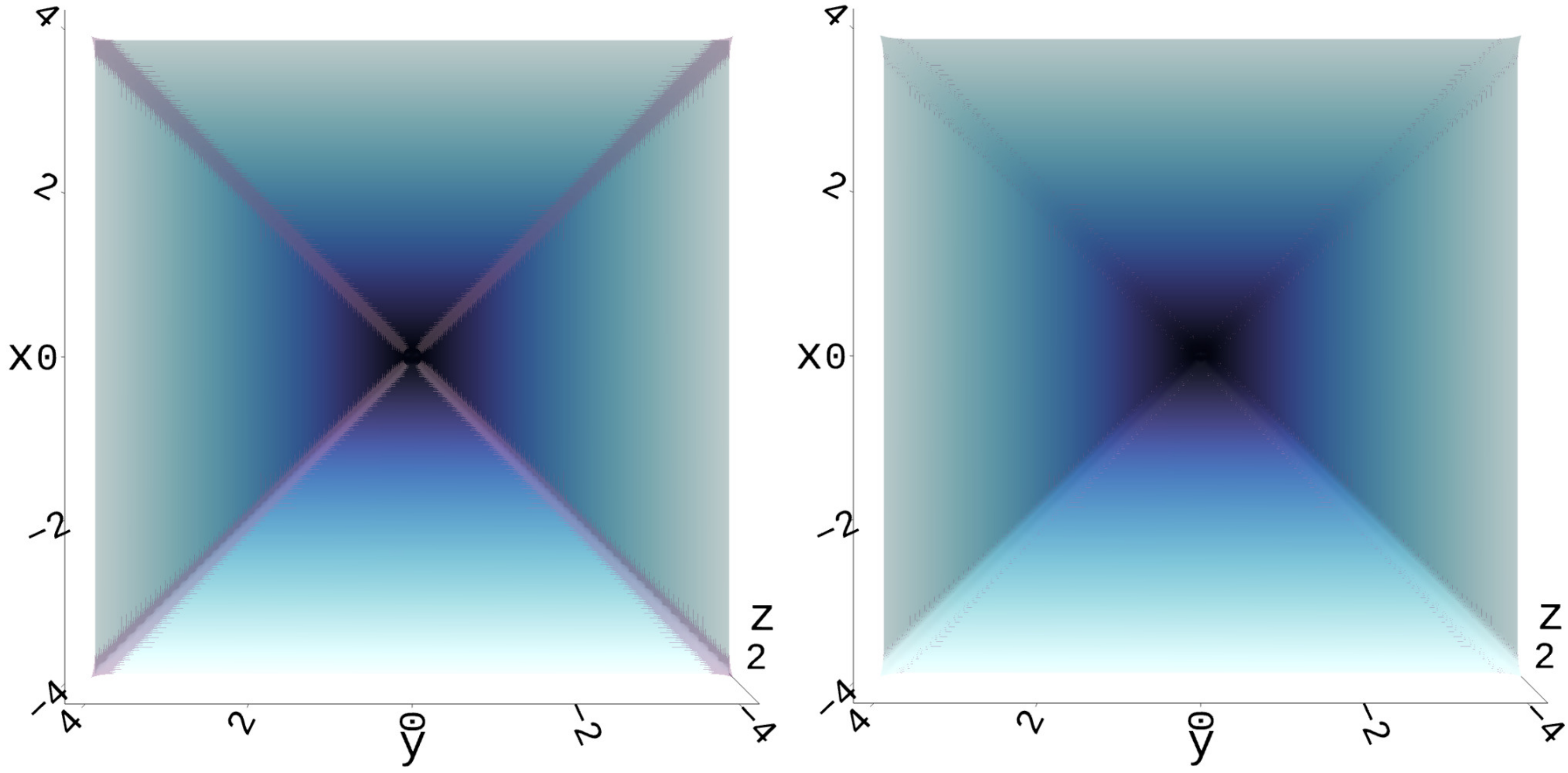}}
\caption{ 
The 2 figures illustrate the value functions.
The left figure displays the approximated value function, with the leaking corners" $\mathcal{L}(t)$—where the values deviate from the ground truth—highlighted in gray.
The right figure shows the value function after applying our correction method, where no ``leaking corners" remain, demonstrating the effectiveness of our method in aligning with the ground truth.
}
\label{fig: 2d_value}
\end{figure}

The value comparison for the final time step is shown in Fig.~\ref{fig: 2d_value}

\subsection{Running Example: 6D Planar Quadrotor}

Consider the 6D Planar Quadrotor:
\begin{equation}
    \dot z = 
    \begin{bmatrix}
        \dot x \\
        \dot y \\
        \dot v_x \\
        \dot v_y \\
        \dot \theta \\
        \dot \omega
    \end{bmatrix} =
    \begin{bmatrix}
        v_x \\
        v_y \\
        - u_T \sin{\theta} \\
        u_T \cos{\theta} - g \\
        \omega \\
        u_\tau
    \end{bmatrix}
\vspace{-1mm}
\end{equation}
The control signal is $u=(u_T, u_\tau)$, with the control constraint $c^1(u_T) = |u_T| \leq \overline u_T = 1 \text{ m/s}$, and $c^2(u_\tau) = |u_\tau| \leq \overline u_\tau=1 \text{ rad/s}^2$.
The state partitions of the system are $x,y,v_x,v_y,\theta,\omega$.

Using the SCSD method, the dynamics could be decomposed into 2 subsystems.
Subsystem 1 has the dynamics:

\vspace{-1mm}
\begin{equation}
    \dot x_1 = (\dot x, \dot v_x, \dot \theta, \dot \omega),
\vspace{-1mm}
\end{equation}
and subsystem 2 has the dynamics:

\vspace{-1mm}
\begin{equation}
    \dot x_2 = (\dot y, \dot v_y, \dot \theta, \dot \omega).
\vspace{-1mm}
\end{equation}

The initial value function for subsystem 1 is $\phi_1(x_1,0)=\ell_1(x_1)= x $, and for subsystem 2, it is $\phi_2(x_2,0)=\ell_2(x_2)=y$.
The initial value function for the full-dimensional system is given by $V(z,0)=\min\{\ell_1(x_1), \ell_2(x_2)\}$.

The grid consists of $21^6$ points.
The $x$ and $y$ dimensions range from $-1.0$ to $4.0$, while the remaining dimensions range from $-2.0$ to $2.0$.
\begin{table}[h]
    \centering
    \caption{6D Accuracy Comparison for One Step}
    \label{table: 6D_OneStep_Accuracy}
    \begin{tabular}{|p{3cm}|>{\centering\arraybackslash}p{2cm}|>{\centering\arraybackslash}p{2cm}|}
        \hline
        \textbf{Metric} & \textbf{Before} & \textbf{After} \\
        \hline
        Number of grid points with different values from the ground truth & $3.89\times 10^{6}$ & 0 \\
        \hline
        Average absolute difference from ground truth & $6.97 \times 10^{-4}$ & $0.0$ \\
        \hline
        Maximum absolute difference from ground truth & $4 \times 10^{-2}$ & $0.0$ \\
        \hline
    \end{tabular}
    \vspace{-5mm}
\end{table}

\begin{table}[h]
  \centering
  \caption{Computation Time and Delta Value for $t$s}
  \label{table: time_comparison_6d}
  \begin{tabular}{|p{1.5cm}|p{1.5cm}|p{4cm}|}
    \hline
    \textbf{$t$ (s)} & \textbf{$\Delta$} & \textbf{Decomposition Time  + Local Updating Time (seconds)} \\
    \hline
    -0.02 & 0.04 & 2.447 + 47.1078 = 49.5548 \\
    \hline
    -0.06 & 0.1212 & 6.769 + 157.3528 = 164.1218 \\
    \hline
    -0.1 & 0.204 & 11.8038 + 250.7859 = 262.5897 \\
    \hline
  \end{tabular}
\end{table}

Table~\ref{table: 6D_OneStep_Accuracy} presents the accuracy comparison for $1$ time step.
The computational time results are summarized in Table~\ref{table: time_comparison_6d}.
This case was previously intractable due to the ``curse of dimensionality."

\begin{figure}[h]
{\includegraphics[width=\linewidth]{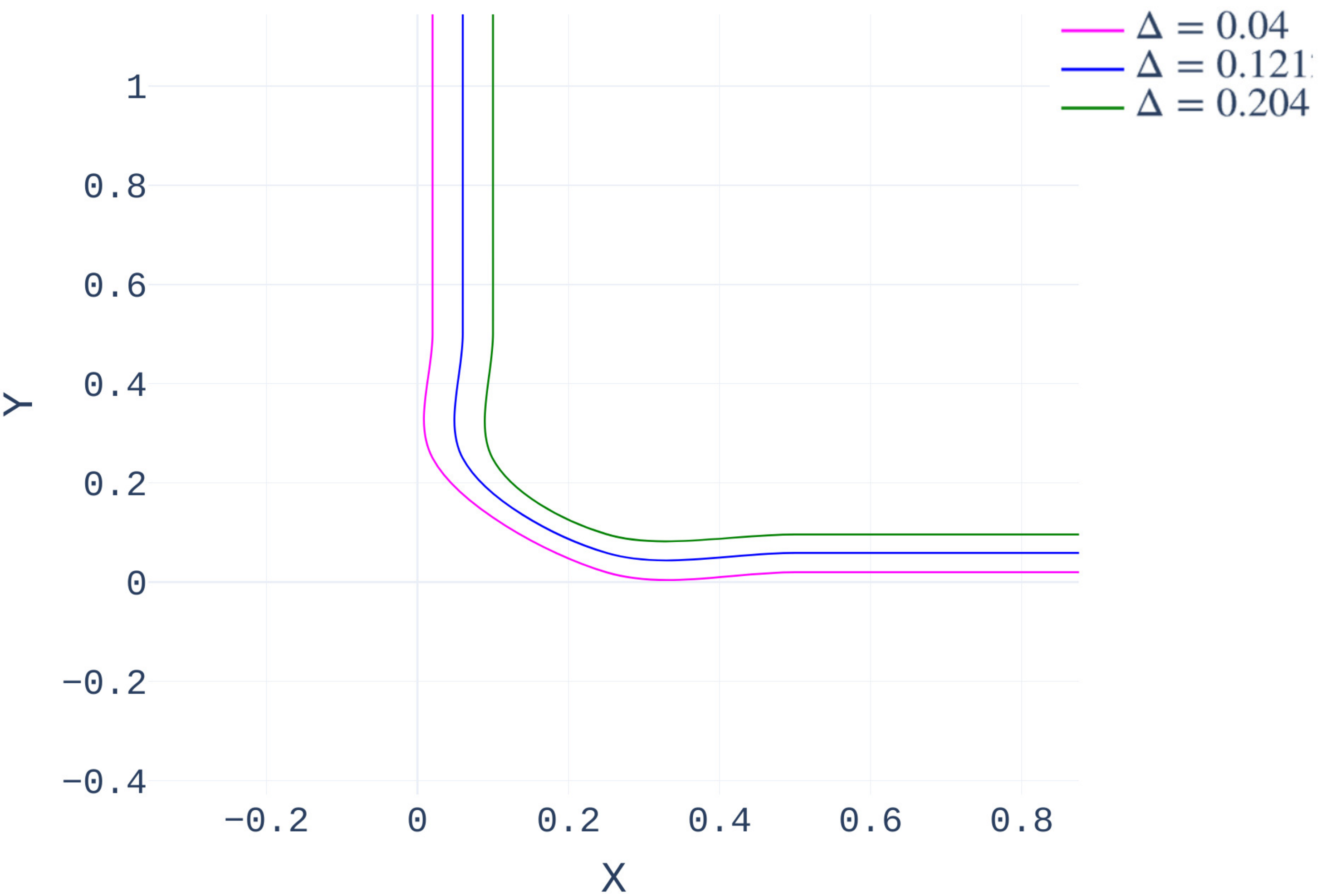}}
\caption{ 
The figure demonstrates the local updating results for the backward computation of $0.02$ seconds with $\Delta=0.04$, $0.06$ seconds with $\Delta=0.1212$ and $0.1$ seconds with $\Delta=0.204$.
The dimensions shown are $x$ and $y$.
We take a slice of $v_x=-1$, $v_y=-1$, $\theta=0$, and $\omega=0.4$.
}
\label{fig: 6d_value}
\end{figure}

The value function slices for different $\Delta$ values at corresponding time steps are shown in Fig.~\ref{fig: 6d_value}.

\section{Conclusion and Future Work}
In this paper, we propose a threshold-based method to detect the leaking corners when low-dimensional control inputs are subject to certain control constraints.
We also introduce a local updating method that ensures accuracy while maintaining computational efficiency. 
The proposed method is validated using a 2D Single Integrator system and a 6D Planar Quadrotor system with the SCSD method.

Future work includes:
(1) Testing the method with other computational acceleration techniques;
(2) Parallelizing the local updating process for faster computation;
(3) Exploring machine learning or other techniques for new value updating methods;

\bibliographystyle{ieeetr}
\bibliography{ref.bib}

\begin{thebibliography}{10}

\bibitem{bansal2017hamilton}
S.~Bansal, M.~Chen, S.~Herbert, and C.~J. Tomlin, ``Hamilton-jacobi
  reachability: A brief overview and recent advances,'' in {\em Conf. on
  Decision and Control}, IEEE, 2017.

\bibitem{wensing2023optimization}
P.~M. Wensing, M.~Posa, Y.~Hu, A.~Escande, N.~Mansard, and A.~Del~Prete,
  ``Optimization-based control for dynamic legged robots,'' {\em IEEE
  Transactions on Robotics}, vol.~40, pp.~43--63, 2023.

\bibitem{natarajan2023human}
M.~Natarajan, E.~Seraj, B.~Altundas, R.~Paleja, S.~Ye, L.~Chen, R.~Jensen,
  K.~C. Chang, and M.~Gombolay, ``Human-robot teaming: grand challenges,'' {\em
  Current Robotics Reports}, vol.~4, no.~3, pp.~81--100, 2023.

\bibitem{hanover2024autonomous}
D.~Hanover, A.~Loquercio, L.~Bauersfeld, A.~Romero, R.~Penicka, Y.~Song,
  G.~Cioffi, E.~Kaufmann, and D.~Scaramuzza, ``Autonomous drone racing: A
  survey,'' {\em IEEE Transactions on Robotics}, 2024.

\bibitem{kong2015dreach}
S.~Kong, S.~Gao, W.~Chen, and E.~Clarke, ``dreach: $\delta$-reachability
  analysis for hybrid systems,'' in {\em Tools and Algorithms for the
  Construction and Analysis of Systems}, 2015.

\bibitem{huang2019reachnn}
C.~Huang, J.~Fan, W.~Li, X.~Chen, and Q.~Zhu, ``Reachnn: Reachability analysis
  of neural-network controlled systems,'' {\em ACM Transactions on Embedded
  Computing Systems (TECS)}, vol.~18, no.~5s, pp.~1--22, 2019.

\bibitem{mitchell2005time}
I.~M. Mitchell, A.~M. Bayen, and C.~J. Tomlin, ``A time-dependent
  hamilton-jacobi formulation of reachable sets for continuous dynamic games,''
  {\em IEEE Trans. on automatic control}, vol.~50, no.~7, 2005.

\bibitem{spaceex2011frehse}
G.~Frehse, C.~Le~Guernic, A.~Donz{\'e}, S.~Cotton, R.~Ray, O.~Lebeltel,
  R.~Ripado, A.~Girard, T.~Dang, and O.~Maler, ``Spaceex: Scalable verification
  of hybrid systems,'' in {\em Comput. Aided Verification}, 2011.

\bibitem{althoff2010computing}
M.~Althoff, O.~Stursberg, and M.~Buss, ``Computing reachable sets of hybrid
  systems using a combination of zonotopes and polytopes,'' {\em Nonlinear
  analysis: hybrid systems}, 2010.

\bibitem{darbon2016algorithms}
J.~Darbon and S.~Osher, ``Algorithms for overcoming the curse of dimensionality
  for certain hamilton--jacobi equations arising in control theory and
  elsewhere,'' {\em Research in the Mathematical Sciences}, vol.~3, no.~1,
  p.~19, 2016.

\bibitem{lee2023efficient}
D.~Lee and C.~J. Tomlin, ``Efficient computation of state-constrained
  reachability problems using hopf--lax formulae,'' {\em IEEE Transactions on
  Automatic Control}, vol.~68, no.~11, pp.~6481--6495, 2023.

\bibitem{sharpless2024conservative}
W.~Sharpless, Y.~T. Chow, and S.~Herbert, ``Conservative linear envelopes for
  nonlinear, high-dimensional, hamilton-jacobi reachability,'' {\em arXiv
  preprint arXiv:2403.14184}, 2024.

\bibitem{kurzhanski2000ellipsoidal}
A.~B. Kurzhanski and P.~Varaiya, ``Ellipsoidal techniques for reachability
  analysis: internal approximation,'' {\em Syst. \& Control Letters}, vol.~41,
  no.~3, 2000.

\bibitem{li2018approx}
M.~Li, P.~N. Mosaad, M.~Fr{\"a}nzle, Z.~She, and B.~Xue, ``Safe over-and
  under-approximation of reachable sets for autonomous dynamical systems,'' in
  {\em Formal Modeling and Analysis of Timed Systems}, 2018.

\bibitem{liu2024efficient}
V.~Liu, C.~Manzie, and P.~M. Dower, ``Efficient value function upper bounds for
  a class of constrained linear time-varying optimal control problems,'' in
  {\em 2024 American Control Conference (ACC)}, pp.~4084--4089, IEEE, 2024.

\bibitem{fisac2019bridging}
J.~F. Fisac, N.~F. Lugovoy, V.~Rubies-Royo, S.~Ghosh, and C.~J. Tomlin,
  ``Bridging hamilton-jacobi safety analysis and reinforcement learning,'' in
  {\em Int. Conf. on Robot. and Automat.}, IEEE, 2019.

\bibitem{bansal2021deepreach}
S.~Bansal and C.~J. Tomlin, ``Deepreach: A deep learning approach to
  high-dimensional reachability,'' in {\em 2021 IEEE International Conference
  on Robotics and Automation (ICRA)}, pp.~1817--1824, IEEE, 2021.

\bibitem{lin2024verification}
A.~Lin and S.~Bansal, ``Verification of neural reachable tubes via scenario
  optimization and conformal prediction,'' in {\em 6th Annual Learning for
  Dynamics \& Control Conference}, pp.~719--731, PMLR, 2024.

\bibitem{ruthotto2020machine}
L.~Ruthotto, S.~J. Osher, W.~Li, L.~Nurbekyan, and S.~W. Fung, ``A machine
  learning framework for solving high-dimensional mean field game and mean
  field control problems,'' {\em Proceedings of the National Academy of
  Sciences}, vol.~117, no.~17, pp.~9183--9193, 2020.

\bibitem{herbert2019warm}
S.~L. Herbert, S.~Bansal, S.~Ghosh, and C.~J. Tomlin, ``Reachability-based
  safety guarantees using efficient initializations,'' in {\em Conf. on
  Decision and Control}, IEEE, 2019.

\bibitem{chow2019algorithm}
Y.~T. Chow, J.~Darbon, S.~Osher, and W.~Yin, ``Algorithm for overcoming the
  curse of dimensionality for state-dependent hamilton-jacobi equations,'' {\em
  Journal of Computational Physics}, vol.~387, pp.~376--409, 2019.

\bibitem{bui2022optimizeddp}
M.~Bui, G.~Giovanis, M.~Chen, and A.~Shriraman, ``Optimizeddp: An efficient,
  user-friendly library for optimal control and dynamic programming,'' {\em
  arXiv preprint arXiv:2204.05520}, 2022.

\bibitem{mitchell2011scalable}
I.~M. Mitchell, ``Scalable calculation of reach sets and tubes for nonlinear
  systems with terminal integrators: a mixed implicit explicit formulation,''
  in {\em Proceedings of the 14th international conference on Hybrid systems:
  computation and control}, pp.~103--112, 2011.

\bibitem{holmes2020reachable}
P.~Holmes, S.~Kousik, B.~Zhang, D.~Raz, C.~Barbalata, M.~Johnson-Roberson, and
  R.~Vasudevan, ``Reachable sets for safe, real-time manipulator trajectory
  design,'' 2020.

\bibitem{chen2017exact}
M.~Chen, S.~Herbert, and C.~J. Tomlin, ``Exact and efficient hamilton-jacobi
  guaranteed safety analysis via system decomposition,'' in {\em Int. Conf. on
  Robot. and Automat.}, IEEE, 2017.

\bibitem{chen2018decomposition}
M.~Chen, S.~L. Herbert, M.~S. Vashishtha, S.~Bansal, and C.~J. Tomlin,
  ``Decomposition of reachable sets and tubes for a class of nonlinear
  systems,'' {\em Trans. on Automatic Control}, vol.~63, no.~11, 2018.

\bibitem{lee2019removing}
D.~Lee, M.~Chen, and C.~J. Tomlin, ``Removing leaking corners to reduce
  dimensionality in hamilton-jacobi reachability,'' in {\em 2019 International
  Conference on Robotics and Automation (ICRA)}, pp.~9320--9326, IEEE, 2019.

\bibitem{he2023efficient}
C.~He, Z.~Gong, M.~Chen, and S.~Herbert, ``Efficient and guaranteed
  hamilton-jacobi reachability via self-contained subsystem decomposition and
  admissible control sets,'' {\em IEEE Control Systems Letters}, 2023.

\bibitem{mitchell2007toolbox}
I.~M. Mitchell {\em et~al.}, ``A toolbox of level set methods,'' {\em UBC
  Department of Computer Science Technical Report TR-2007-11}, 2007.

\bibitem{evans2022partial}
L.~C. Evans, {\em Partial differential equations}, vol.~19.
\newblock American Mathematical Society, 2022.

\end{thebibliography}

\end{document}